Two-dimensional gel electrophoresis in proteomics: past, present and future

Thierry Rabilloud[1,2], Mireille Chevallet[2], Sylvie Luche[2,3] , Cécile Lelong [2,3]

CNRS UMR5092, Biochemistry and Biophysics of Integrated Systems, CEA Grenoble, iRTSV/BSBBSI, 17 rue des martyrs, F-38054 GRENOBLE CEDEX 9

CEA-DSV/iRTSV/LBBSI, Biophysique et Biochimie des Systèmes Intégrés, CEA-Grenoble, 17 rue des martyrs, F-38054 GRENOBLE CEDEX 9, France

Université Joseph Fourier, UMR CNRS-CEA-UJF 5092, CEA-Grenoble, 17 rue des martyrs, F-38054 GRENOBLE CEDEX 9

Correspondence to

Thierry Rabilloud, iRTSV/BBSI

CEA-Grenoble, 17 rue des martyrs,

F-38054 GRENOBLE CEDEX 9

Tel (33)-4-38-78-32-12

Fax (33)-4-38-78-44-99

e-mail: Thierry.Rabilloud@ cea.fr

Abstract

Two-dimensional gel electrophoresis has been instrumental in the birth and developments of proteomics, although it is no longer the exclusive separation tool used in the field of proteomics. In this review, a historical perspective is made, starting from the days where two-dimensional gels were used and the word proteomics did not even exist. The events that have led to the birth of proteomics are also recalled, ending with a description of the limitations of two-dimensional gels in proteomics. However, the advantages of two-dimensional gels are also mentioned leading to a critical description of how and when to use them best in a proteomics approach. Taking support of theses advantages (robustness, resolution, and ability to separate entire, intact proteins), possible future applications of this technique in proteomics are also briefly mentioned.

1. From mute star maps to protein buzz: Two-dimensional electrophoresis and the birth of proteomics

SDS electrophoresis in its modern form was introduced in the early 70's [1] and soon became very widely used by protein biochemists. However, analysis of complex cellular extracts by this method made obvious the fact that the resolution was far from being sufficient to separate the many protein components of such extracts. To increase the resolution to a sufficient extent, it was necessary to couple two independent separations. At that time (and this is still the case) it was obvious that the separation best able to complement SDS electrophoresis was isoelectric focusing, and this was the coupling that researchers in protein separation tried to achieve. One of the first reports of successful two-dimensional electrophoresis coupling IEF to SDS PAGE was published in 1974 [2] but go relatively unnoticed by the community, probably because of two features. The first one was the inclusion of the sample within the IEF tube gel, which is a quite efficient but slightly cumbersome process. The second was the detection method chosen, i.e. the classical Coomassie blue staining in alcohol-acid mixtures. The poor sensitivity of the method led to nice but rather "empty" 2D images, and therefore not that impressive. The situation changed dramatically the year after, with the seminal paper of PH O'Farrell [3]. This paper coupled better technical choices (loading of samples on top of IEF gels), a very detailed description of the protocol, and detection by autoradiography, showing at one hundreds of protein spots on a single gel. Many researchers got very impressed by the results and underlying power of the method, so that 2D electrophoresis spread quite widely and quickly around the world. This is clearly shown by the citation index of the paper, gathering more than 1500 citations from 1976 to 1980. As soon as that time (late 70's very early 80's), the first applications of what was to become proteomics took off, with applications to cell biology [4, 5] and clinical biology [6], with the first ancestor of the human proteome project [7].

However, it must absolutely be stressed that protein identification at that time was quite an ordeal. The main methods were comigration with purified proteins [8] and blotting with antibodies [9, 10]. Thus, it was possible to say where was protein X (purified or with an available antibody) on the gels, but it was very difficult to know who was spot Y. For example, it took several years to know that cyclin/PCNA, identified as a cell cycle dependent protein [11] was indeed a subunit of DNA polymerase delta [12].
In the absence of readily-available protein identification tools, lot of attention was brought to the use of quantitative protein patterns as identifiers of cellular states via correlation analyses. It is often forgotten that statistical analysis in proteomics did not start a few years ago by getting inspired from transcriptomics, but long before the word proteomics was even coined, and in fact some 20-25 years ago. Indeed, computer analysis included both gel image analysis [13-16], but also strong data analysis with multivariate methods [17-22].
However, it was not by chance that one of the first 2D gel analysis system was name TYCHO [14], after the name of the famous astronomer (and alchemist) Tycho Brahe, as 2D maps were much like silent star maps for more than ten years (1977 to 1987).
In addition to this dearly-resented lack of universal protein identification system, 2D

electrophoresis was plagued at that time with major reproducibility problems, and was much more an art than everything else. New practitioners of the field may not realise that the IEF gels were soft unsupported tube gels (typically 4% acrylamide) that were to be extruded from the glass tube containing them for the first dimension in order to be loaded on the second dimension slab gel. This was quite a difficult process, and a lot of non-linear deformation of the tube gel took place there. Ending with lengths variations of 10-20% was not uncommon, and this was highly detrimental to spot positional reproducibility. Moreover, IEF with carrier ampholytes is by itself a process that must be carefully controlled to achieve reproducibility. Carrier ampholytes-based pH gradients are not really stable, and prone to cathodic drift, i.e. progressive loss of the basic portion of the gradient, so that the migration profile (in Volts and Volt.hours) had to be carefully controlled. Moreover, the fine profile of the pH gradient is dependent on the precise composition of the carrier ampholytes, and this varies from batch to batch, leading to long-term irreproducibility of the IEF separations.

Despite these difficulties, exquisite experiments could be devised, as the demonstration of the expression of several actin forms in a single cell [23]. In addition, important dimensions of our current proteomic knowledge root from that period, such as the number of proteins present in a mammalian cell [24] or the knowledge that serum proteins are extensively modified [8].

The best was to come in the last 80's, with two key events for the development of two-dimensional electrophoresis in biology. The first key event was the introduction of immobilized pH gradients in the first dimension. By construction, immobilized pH gradients eliminate the reproducibility problems associated with carrier ampholytes-based pH gradients. However, initial attempts to adapt them for 2D electrophoresis, either in the tube gel mode [25] or via a gel slab [26] were not very successful, and the final solution, i.e. the now classical gel strips, was introduced later [27]. Further work was necessary to ensure good interfacing of the IPG strips with the SDS slab gel [28], but a stable protocol was soon available for the community [29, 30].
With the establishment of iPGs, 2D electrophoresis ceased to be a difficult art to convert only into good craftsmanship. There was a real quantum leap in reproducibility, going up to interlaboratory studies with a positive outcome [31, 32], something that was just unbelievable a few years before.

The real change of paradigm, however, came with the long-waited interface of 2D gels with protein microanalysis techniques. This was some years before mass spectrometry entered the scene, and obtained through the progress of Edman sequencing. The late 80's saw a blossoming of microsequencing applied to gel-separated proteins, either directly N-terminal sequencing [33], easy to perform and sensitive (only 1-2 picomoles required) but requiring a free N-terminus, something not often available in eukaryotic proteins. To bypass this limitation, internal sequencing protocols were developed, either from blots [34], [35] or directly from stained gels [36, 37]. This process of internal sequencing was more sample-consuming (around 50-100 picomoles required) and technically difficult, requiring microseparation of peptides on narrow-bore HPLC, collection of individual peptides by UV trace and sequencing of these individual peptides. In many studies, obtaining 50 picomoles of the protein(s) of interest was not an easy task, but was still achievable by using the high loading capacity of IPG gels [38-40], and devices to pool the proteins extracted from several gel spots [41, 42].

2. The take-off of proteomics

This possibility to be able to analyze spot of interest coming from 2D gels was a crucial change of paradigm, and the real start of proteomics. At those times where no complete genome was published yet, Edman sequencing provided enough information to look for homologs, or to devise oligonucleotides for screening DNA libraries. It was also the starting of brief golden age of 2D gel images databases [43-47]. To be useful to the community, 2D gel image databases must contain enough information, and not be just descriptive. Therefore, generic methods for obtaining information on proteins must be available. However, if these methods become easy and widespread, every laboratory can make its own identifications, and the added value of a 2D gel image database decreases. This is why the development and use of 2D gel databases culminated at this period, where protein identification was feasible but not very easy, i.e. in the early 90's.

Around the middle of the 90's, a quantum leap in protein identification arrived with the introduction of mass spectrometry-based methods. The first methods were based on peptide mass fingerprinting [48-51]. In order to work, these methods need a pure protein, or at least a mixture where one protein dominates by far the other ones in a quantitative point of view, something that 2D gels provide rather easily, especially when narrow pH ranges (here again available through the IPG technology) are used [52]. Compared to Edman sequencing, peptide mass fingerprinting brings less information on sequences, but some on putative post-translational modifications, and is by far more productive and less protein consuming. Even at these rather early days of peptide mass fingerprinting, the demand was in the low picomole range, i.e. at least one order of magnitude below the one of internal Edman sequencing, and the analysis time was counted in hours of work, not days, which allowed for a much improved productivity in proteomics.

Even though the coining of the word "proteomics" as simultaneous with the onset of practicable mass spectrometry methods in the field, it was not the true birth of proteomics. However, it was its first real blossoming, and numerous laboratories used this combination of 2D gel electrophoresis and peptide mass fingerprinting to carry out proteomics work in various areas of biology, and also for feeding databases [53-58] . At that time, many researcher, although conscious that some challenges remained [59] , were very confident that this combination of tools would be able to resolve complete proteomes [60], but it soon appeared that this would not be the case.

3. Touching the limits of 2D electrophoresis, and the creation of alternate methods.

Because of this intensive and worldwide effort using 2D electrophoresis and mass spectrometry as the core tool, many data were accumulated and analyzed, and it soon became obvious that it was always the same types of proteins that were found again and again, and the same types that were always missing, i.e. the low abundance and the hydrophobic proteins [61, 62].

In order to improve the resolution of hydrophobic proteins, many efforts were devoted to improve protein solubilization under the conditions prevailing in the IEF dimension. This included changes in the chaotropes used in IEF [63] and also in the detergents used for this step. Better solubilization for membrane proteins in 2D gels had been identified as a problem since the very beginning of the technique [64], and some improvements had been proposed over the years [65-67]. However, the increasing evidence of the problem prompted several groups to use various types of IEF-compatible detergents to alleviate this problem [68-71]. There was some success, i.e. demonstration of membrane proteins in 2D gels. However, it is fair to say that the problem of hydrophobic, and especially membrane proteins on 2D gels is largely unsolved [72-74]. Moreover, as this problem is clearly correlated with the IEF dimension and to the chemical conditions prevailing at this step (low ionic strength, no ionic detergents) [75], it is quite obvious that this will remain a built-in problem for all IEF-containing 2D electrophoresis systems. Thus, for analyzing membrane proteins, IEF-free separation systems were required. However, as the IEF/SDS combination is the only one which separation power matches the complexity of proteomes, this meant in turn that IEF-free separations would only lead to protein mixtures in the fractions arising from the separation, so that identification techniques able to cope with mixtures must be used. Fortunately enough, this had been worked for some years by developing the use of tandem mass spectrometry [76-80]. Thus, efficient IEF-free proteomic schemes could be devised, using peptide separations only, as in the shotgun approach [81, 82], or a combination of protein SDS electrophoresis followed by a peptide separation [83] [84]. In fact these setups proved able to analyze the hydrophobic proteins refractory to classical 2D gels.

The problem of the low abundance proteins is also more acute in 2D gel-based proteomics than in these other setups for several reasons.
The first one is the simple fact that 2D gel-based proteomics is the only proteomic setup in which there is a readout before mass spectrometry. In other words, 2D gels are not gridded blindly with each and every piece of gel sent out for digestion and MS analysis. Thus, their performance, and especially their ability to detect and quantify low abundance proteins, is dependent on a protein visualization method after electrophoresis. Although modern high-sensitivity detection techniques (silver stain and fluorescence) are able to operate in the low nanogram range, with fluorescence-based detection being linear over several orders of magnitude, there is no detection method able to cope with the enormous dynamic range (i.e. quantitative ratio between the rarest protein expressed in a sample and the most abundant one) present in most biological samples [85] and even worse in biological fluids [86].

The second reason, going along the same line, is the distribution of the protein abundances, i.e. the proportion of proteins expressed at a rather high concentration vs. the proportion of proteins expressed at low concentrations. Such figures are much more difficult to obtain, but have been thoroughly investigated in yeast [87], and the results are appalling. Of the almost 4000 gene products that could be quantified, 130 gene products account for half of the protein content in yeast cells. The 10% most-expressed gene products account for 75% of the protein content, and the 2/3 less-expressed proteins account for only 10% of the protein content. Thus, being limited in the detection of low abundance proteins also means leaving a large proportion of cellular proteins out of he analysis.

There is an obvious countermeasure to alleviate this problem, which would be to load more sample, taking advantage of the huge capacity of 2D gels [38-40], and thus bringing many more proteins above the detection limit. But here the third limitation comes to play to keep the usefulness of this high-loading approach in check. This limitation is linked to gel crowding and is linked to the protein abundance distribution. As 2D gels are able to resolve many modified forms of the proteins, the modified forms of the high abundance proteins will occupy precious separation space on 2D gels [59]. Consequently, loading much more sample will result mainly in completely obscuring some zones, leading in no added performance for the detection of the low abundance proteins in these zones [24]. A possible retort would be to use giant gels with a much increased resolution and capacity [88], but this technology is rather difficult to use due mainly to the fragility of the giant gels.
As this problem is linked mainly to the ability of 2D gels to separate modified forms, mainly through the IEF dimension, it is easy to understand that this problem is much more severe in this configuration than in IEF-free methods.

Last but not least, it should be stressed that 2D gels have a rather moderate overall yield [89], with important losses at the IEF stage and during equilibration between the IEF and SDS dimensions.

However, it shall not be derived from this section that 2D electrophoresis is the only proteomic method with a poor yield and unable to deal with low abundance proteins. All proteomics setups bump into the rare proteins problem, as exemplified by the fact that no proteomic method has proven able to analyze comprehensively a complex biological sample (even a bacterium). Apart from the problem of hydrophobic proteins, which is really a built-in weakness of 2D gels, 2D gels just perform differently from other proteomic setups. Although it is true that non-IEF proteomic setups perform better for low abundance proteins, it should be noted that on any sample type, 2D gels are still able to analyze protein that escape analysis by other methods, as shown by the examples of nucleolus proteomics [90, 91], or by the detection of cytokines in secreted proteins [92].

Instead, it should be analyzed that the weaknesses of 2D gels are now well-known because this is the oldest technique, used by a wide community of researchers who have done their homework and investigated honestly the limits of their tools. Because of their more recent introduction, other proteomic setups have not been subjected to the same testing and their limits are not as well- and widely-known yet. And despite these now well-known limitations, 2D gels still offer may attractive features that can be very useful in proteomics-based research.

4. Using the strengths of 2D gels in modern proteomic research

In this section, we will try to sort out the major strengths of 2D gels and to exemplify how they can be used for the benefit of proteomic research.

4.1 Robustness and technical confidence

In fact, because of the hindsight developed by the proteomics community on 2D gels, the drawbacks of 2D gels are well-known, but their advantages are also very well-known. One of the major advantages of 2D gels in proteomics lies in the robustness of the technique. As mentioned earlier, this robustness has been tested thoroughly [93], even in interlaboratory comparisons [31][32], and the influence of the various parameters on the intralaboratory reproducibility have also been investigated

[94]. In fact, the most critical variable nowadays are no longer in the 2D gel process per se, but rather upstream and downstream, i.e. sample preparation and image production and analysis.

As to image production, most often made now by protein stains, it has made enormous gains in reproducibility over the more than 30 years period in which scientists have run 2D gels. Modern detection methods, such as fluorescent stains [95, 96] colloidal Coomassie blue [97], and even modern silver staining [98], where development goes to an end-point, all show a modal coefficient of variation (CV) of ca. 20% (including the variation of the 2D gel process). In addition, most of these detection methods are fairly mass-spectrometry compatible [99], so that sequence coverages in the range of 25-50% are very common. This is due not only to the performances of the staining procedures themselves, but also to the fact the 2D electrophoresis produces a good separation and a concentration process of the proteins of interest into the spots, so that the mass spectrometry process is focused on the protein of interest and not polluted by contaminating proteins (if good laboratory practices excluding the airborne contaminants are used, of course). Classical silver staining methods do not perform as well as to compatibility with mass spectrometry [100], although specialized variants show better compatibility with mass spectrometry [101]. This problem has been shown to be correlated with the use of formaldehyde as the staining-developing agent [102], and consequently completely formaldehyde-free silver staining methods have been introduced [103].
Moreover, the technical variability has decreased with the use of multiplexed electrophoresis [104]. In this system, where different samples are differentially labelled with different fluorophores and then mixed before migration on a single 2D gel, it has been shown that reproducibility and precision were greatly increased compared to the standard system [105].

Thus, the main remaining factor for variability lies in the sample itself. Moreover, 2D electrophoresis, (and especially the IEF dimension) is very sensitive to many interfering compounds present in most biological samples, so that sample preparation must be adapted to each sample type.

However, the main source of variability is still the biological sample itself. This holds especially true at both ends of biological complexity, i.e. for mammalian or plant tissues or fluids on the one hand, and for bacteria on the other hands. In the former case (plants and mammals) the genetic heterogeneity and the poor experimental control of the physiological states are the major sources of variability. In the latter case (bacteria), the metabolic flexibility is the problem, as minor differences in the culture conditions will result in metabolic adaptation and thus to proteome changes that are easily detected [106]. However, when this parameter is well-controlled, the combination of the limited complexity of bacteria [85] with the resources of bacterial genetics makes the 2D gel/MS a very successful tool for studying bacteria at a proteomic scale for very various aspects [57, 107-112].

4.2. Parallelism and statistical confidence

Parallelism of 2D gels is an often overlooked, but quite important aspect in proteomics. It has often been stated that running 2D gels is a poorly automatizable process, needing 3-4 days of highly-qualified staff time. This is absolutely true, but it is no less true that the same time and moderately higher effort are required to run not one but twelve or even twenty gels. Due to the poor reproducibility of 2D gels in their early days, the pioneers of the field foresaw that parallel running of 2D gels was a critical issue to succeed [113, 114]. This trend has survived over the years, and parallel running of gels is a very common practice, with some obvious benefits. The first one is that the criticism of poor confidence that can be raised toward more recent methods [115] does not apply to 2D gel-based proteomics, where the level of requirement from journals is much higher [116, 117]. Parallel running of biological replicates, i.e. the only way to gain statistical confidence in complex analyses, is routine business in 2D gel-based proteomics, and this is not always the case in other flavors of proteomic analyses.

Moreover, this possibility of making easily multiple comparisons, and not only binary ones, is of high value when complex, non-binary, biological situations must be handled, as in the example of plant-bacteria symbiosis [118]. It is also of high value when very large series of samples must be handled, as in toxicological studies [119-121].

4.3. Adequate use of 2D gels in classical proteomic research

When combining the above-mentioned advantages with the sensitivity to dynamic range issues exposed in section 3, it comes to attention that 2D gels offer a reliable analysis but are limited in the available range of proteins that can be analyzed at a single time, and thus in the optimal complexity of the sample.

This explains several trends that can be observed from the literature, and can be summarized rather easily as "the lower the complexity, the better the performance".

Consequently, as stated earlier, 2D gel-based proteomics performs very well on bacteria and also, although at a lesser extent, on lower eukaryots such as yeast, where exquisite cell regulation experiments have been published [122-124]. However, when going up in genetic complexity and protein expression dynamic ranges [85], the figures of merit decrease rapidly, and 2D gel-based proteomics analyzes only a limited fraction of abundant and soluble proteins, leading to restrictions in the relevance of the observed events. This holds true for mammalian cells [125, 126], but also for biological fluids [127]. This does not mean, however, that 2D gel-based proteomics shall not be applied to this type of sample. When the biological question of interest can be answered, at least partially, within the type of proteins amenable to analysis by 2D gels, the operational advantages of 2D gels (robustness and reproducibility) operate at full strength, and allow to obtain very relevant results as shown on mammalian cells (e.g. [128-130]), but also on plants [131].

However, the best way to use 2D gel-based proteomics is either to focus the question, as exemplified on cell biology-oriented proteomics [132-135], or to lower the complexity of the sample.
In the field of clinical proteomics, this means not to analyze complete serum or plasma, but biological fluids of lower complexity, such as cerebrospinal fluid, which led to a very old success story of proteomics in this field [136][137], but has also been used in more recent research [138]. In this field of clinical proteomics, other fluids can be used, such as tumor interstitial fluid [139], but also cellular extracts, with further confirmation of the putative marker by serum dosage [140, 141].

In the field of cell biology, lowering the complexity of the sample often means analyzing a cellular subfraction such as an organelle. This approach has shown to be efficient on several organelles, such as mitochondria [142-145], but one of the nicest examples might be found in the change of paradigm in the field of phagocytosis [146] and involving the ER, following the observation of the ER protein flotillin in the proteome of phagosomes [147]. Such benefits can be found with other, quite different types of samples, such as detergent-resistant domains [148] or plant cell walls [149].

Thus, the take-home message of this section is to adapt the complexity of the sample to the resolving power of the method, something that sounds trivial in the field of analysis but has not been made that often in the field of proteomics, and especially with 2D gel-based proteomics. In the field of classical proteomics, where proteins presence or change in amount is the parameter of interest, this is the price to pay to take advantage of the robustness and reproducibility of 2D gels. However, 2D gels also offer other, more subtle advantages that can be of even greater value for more precise studies.

4.4. Interface with other biochemical methods (e.g. antibody-based)

One of these "hidden" advantages is the easy and efficient interfacing of 2D gels with other biochemical techniques, and especially those based on antibodies. In fact, antibodies can be used in two formats in biochemistry. They can be used as analytical reagents, to detect and quantify the antigen, or they can be used as micropreparative reagents, to purify the corresponding antigen from a complex sample, a process called immunoaffinity or immunoprecipitation, and relying on the immobilization (direct or indirect) of the antibody on a solid support. Although of utmost interest in the field of proteomics, this immunopurification process is plagued by severe artefacts. These artefacts are mostly due to the spurious binding of unwanted proteins on other regions of the antibody than on the antigen-biding site, or directly on the bead itself [150], or to the contamination of the eluate with antibody-derived protein fragments, as the quantitative yield of this procedure is usually very low.
2D gels, however, can be used with antibodies in the other format, i.e. purely for detection purposes, in the classical blotting setup. In this scheme, the immunoblotting step is used to assign the protein(s) of interest on a reference 2D map, and a subsequent 2D gel is run to perform the identification and characterization of the protein(s) of interest. The overall value of this multistep process relies very heavily on three key parameters:

i) the specificity of the immunodetection. In this frame, it must be stressed that tricks that are completely impossible to use in a preparative scheme, such as competing proteins to increase the specificity of the process, can easily be used in a purely detection-centered process

ii) the reproducibility of 2D gels, to ensure the smooth transposition from gels to blots and vice versa, and even more

iii) the resolving power, as the major caveat would be comigrating proteins, where the antibody would identify the minor component and the subsequent proteomic analysis would identify the major, unrelated component.

Thus, when this scheme is used, its specificity should be assessed, ideally, and when possible, by proteomic identification of the determinant recognized by the antibody (when known), or at least via the use of zoom gels (narrow pH gradients), where the probability of comigration falls.

This blotting-using scheme has been used from the infancy of 2D gels [9]. While it has lost interest as a general protein identification scheme, it still has very valuable application in two principal fields.

The first one is the clinical field, where it can be of interest to know what antigens from a pathogen (including cancer cells) are recognized by the immune system of the patients. In this case, the 2D gel of the pathogen is probed with patients sera, and the recognized proteins are then identified by classical, mass spectrometry-based proteomics. Several examples involving bacteria [151, 152], fungi [153] or cancer cells [154]

The second one is the field of post-translational modifications, using the fact that more and more antibodies are available to detect, directly or indirectly, modified amino acids. The main examples of modified amino acids detected by this scheme are phosphotyrosine (e.g. in [155-157]) nitrotyrosine (e.g. in [158-160] ), but also other oxidative stress-induced modifications such a citrullination [161], protein carbonylation (e.g. in [162-164]), hydroxynonenal adducts [165] or changes in the thiol oxidation, using an immunodetectable, thiol-labelling agent [166]

Thus, this coupling of immunoblotting with the resolution of 2D gels and the analytical power of mass spectrometry allows to perform very efficient analyses of modified proteins, e.g. in a pathological context. However, this scheme is dependent on the availability of good antibodies against modified aminoacids, and only a supervised detection of modified amino acids is possible in this setup.

4.4. Antibody-free analysis of PTM

Besides the use of antibodies, 2D electrophoresis offers the possibility to detect modified form of proteins, mainly on the basis of changes in their pI. The first modification that has been tracked by this process is probably phosphorylation. To give a single example among a numerous literature, protein phosphorylation has been studied in GH cells [167], leading to the discovery of stathmin [168]. In this example, 2D gels have shown their ability to separate and quantify the various phosphorylated states of this protein [169].
In the case of phosphorylation, use of 2D gels relies on the use of radioisotopes, as exemplified above or in [170], or nowadays on enrichment procedures [171] or on the use of selective stains [172].
In another flavor of supervised but antibody-free PTM detection, the study of protein glutathionylation deserves mention through its elegant combination of radiolobelling, reducing vs. non-reducing 2D gels and mass spectrometry-based protein identification [173].

2D gels, however, offer another way of detecting post-translational modifications, in an unsupervised way, by taking advantage of the change in pI induced by many modifications (e.g. phosphorylation). In this scheme, the modified form is first visualized on 2D gels as an extra spot not migrating together with the principal spot, and then the name of the game is to identify the modified peptide(s) by classical mass spectrometry-based analysis. Quite often, the modified peptides are difficult to identify smoothly in mass spectrometry, either because the modification confers an extra negative charge (e.g. phosphorylation), removes a cleavage site (e.g. lysine acylation or methylation) resulting in extra-long peptides, or changes the hydrophobicity/hydrophilicity of the peptide (e.g. glycosylation) resulting in impaired extraction or separation of the peptide. In such cases, the ability of 2D gels to separate the modified form from the bulk of the unmodified one, coupled with their relatively high loading power, are decisive advantages to succeed. Due to these difficulties, there are not that many examples of such approaches in the literature. One can cite, however, the study of adiponection modifications [174], the evidencing of protein deamidation [175], a modification that is quite difficult to detect (low change in mass), for which no role is know, and that is likely to be more widespread than commonly thought, and cysteine overoxidation, especially in the case of peroxiredoxins [176]. This latter example is typical of the difficulties encountered in this approach. The oxidative stress-related modification of peroxiredoxins had been described earlier [177], but without the characterization of the modification, which required 50 picomoles of oxidized peroxiredoxin, due to the length of the peptide (more than 3000 Da) and decrease in flying ability due to the cysteine oxidation. This required to use the full loading capacity of 2D gels to meet such requirements (more than 5 mg of total proteins loaded on the gels). Interestingly enough, when things go to this level of difficulty, the certainty of the modification and of its importance, brought by the 2D gels, is a key argument to go on with the study.

4.5. Analysis of complete proteins

The above examples show how two key features of 2D gels may be used in modern proteomics.The first one is, as mentioned, the ability of 2D gels to serve as a micropreparative tool, a feature that is still used even in complex schemes [178].
The second, which is implicit but need to be strongly emphasized, is the unique ability of 2D gels to be a high-resolution method separating complete proteins, with all their modifications. This is of key importance in studies in which the filiation of peptides (i.e. which peptides belong to which protein) is an important feature.
Interestingly enough, one of the key applications of this feature is to find proteins that are degraded under certain conditions, e.g. apoptosis. In this scheme, a native cell extract is treated (or not) with the protease of interest. The two resulting lysates are separated by 2D gels, the differences are recorded and then characterized by mass spectrometry [179, 180]. This approach has been applied to several proteases involved in apoptosis, such as granzyme B [179], caspase 3 [180], caspase 6 [181] and caspase 7 [182]. Up to now, mostly the decrease of the target proteins has been analyzed, but the often-overlooked ability of 2D gels to separate rather low molecular weight proteins (e.g. in [183]) may allow to investigate the cleavage fragments and thus the cleavage sites.

However, this unique ability to separate complete proteins with a high resolution can be used for other purposes than studying protein degradation. It can be used to prepare a well-defined, pure protein for other purposes such as antibody production, [184], or to study the landscape of protein modifications on selected proteins [185-187].

5. Concluding remarks. What future for 2D gels in proteomics

From all of the above, it is tempting to try to predict the future uses of 2D electrophoresis in the future, although this prediction exercise is always difficult and risky. It can be analyzed, however, that in the present proteomics landscape, 2D electrophoresis has two main drawbacks and three main advantages. The two main drawbacks are its very low efficiency (to say the least), in the analysis of hydrophobic proteins, and its high sensitivity to the dynamic range and quantitative distribution issues. The three main advantages are its robustness, its parallelism, and its unique ability to analyze complete proteins at high resolution. When combining all these features, it becomes obvious that 2D gels will deliver their technical advantages when samples with a limited range of protein expression will be used (e.g. bacteria or cellular subfractions), in order to go beyond the classical "déjà vu" [125, 126], unless this class of proteins is of interest in the biological context (e.g. study of cellular stress).
However, one of the key areas where 2D gels should deliver in the future will be the study of modification landscapes, i.e. how protein modifications combine (or exclude mutually) to modulate protein activity in cells. So, as stated earlier [188], when we will have to go to the details of protein functions, it can be predicted that 2D gels will again deliver their full power.

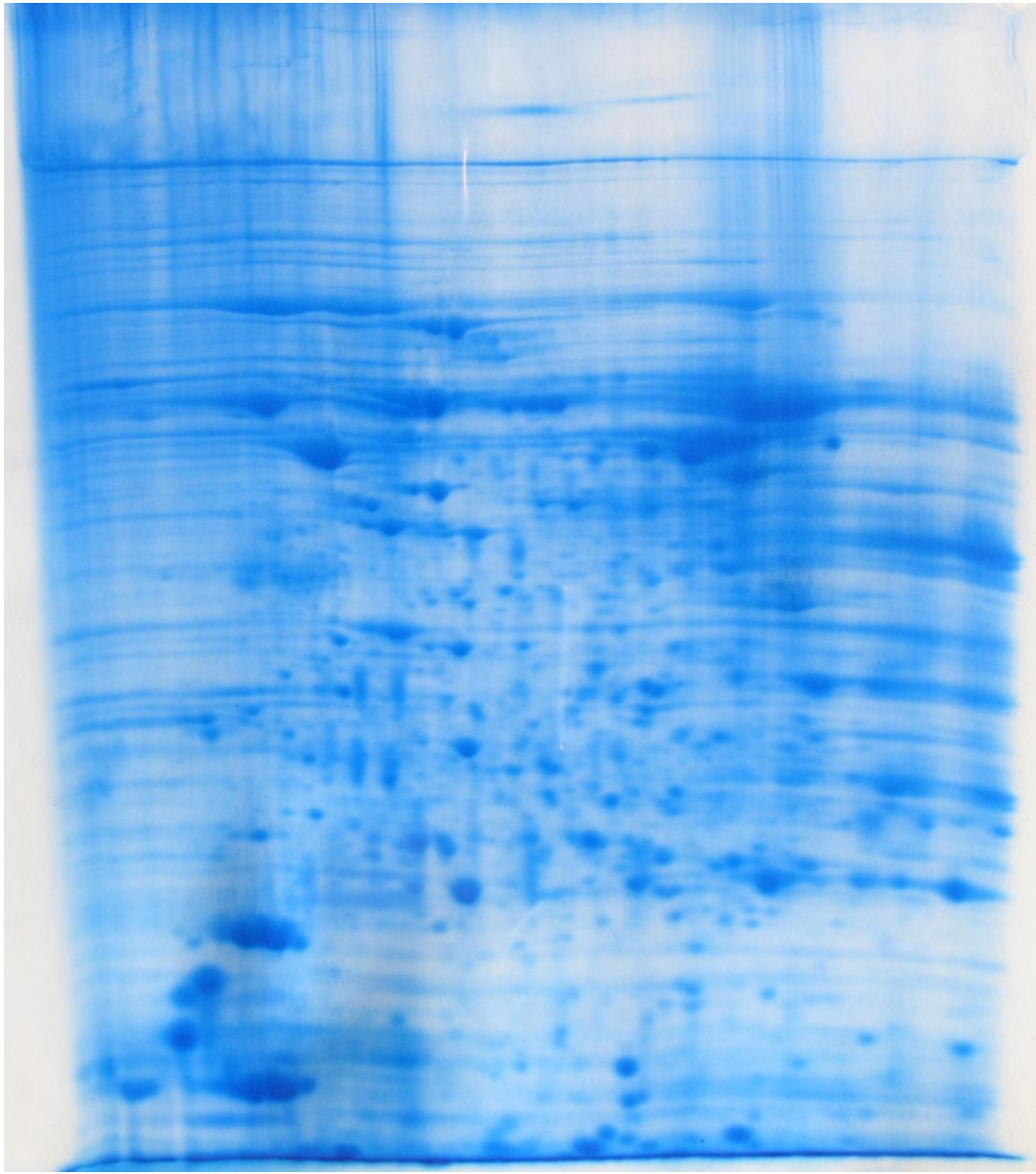

Figure 1. Micropreparative capacities of 2D gels
15 mg of mouse liver mitochondrial proteins were loaded by in -gel rehydration on a linear 4-8 pH gradient, 6mm-wide IPG strip. 1st dimension migration: 65 kVh. Second dimension: 1.5 mm thick, 10% SDS gel. Staining by colloidal Coomassie Blue

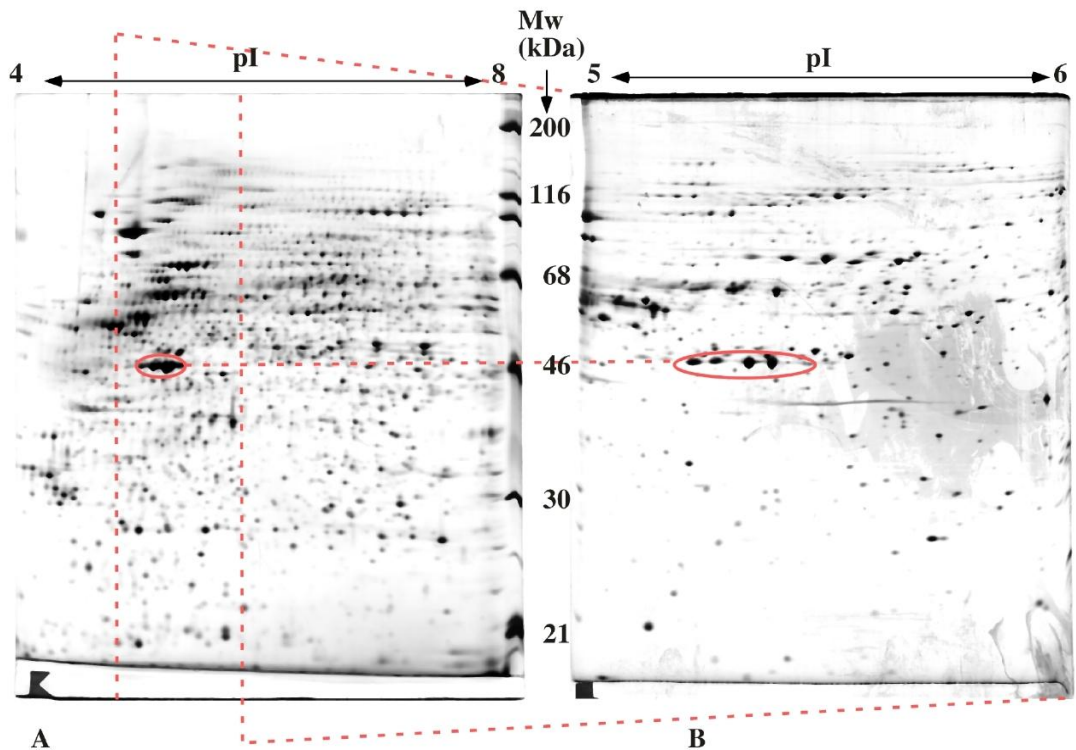

Figure 2: increase in resolution with narrow pH gradients
100 micrograms of protein extracted from HeLa cells were separated by 2D gels (immobilized 16 cm-long pH gradient in the first dimension, 10% continuous acrylamide gel in the second dimension, detection with silver staining). A: separation in the first dimension on a linear pH gradient ranging from 4 to 8. B: separation in the first dimension on a linear pH gradient ranging from 5 to 6. In both cases, IPG gls were migrated for 70 kVh
The oval zone represents the actin spots, almost fused in the 4-8 gradient and individually separated in the 5-6 gradient.

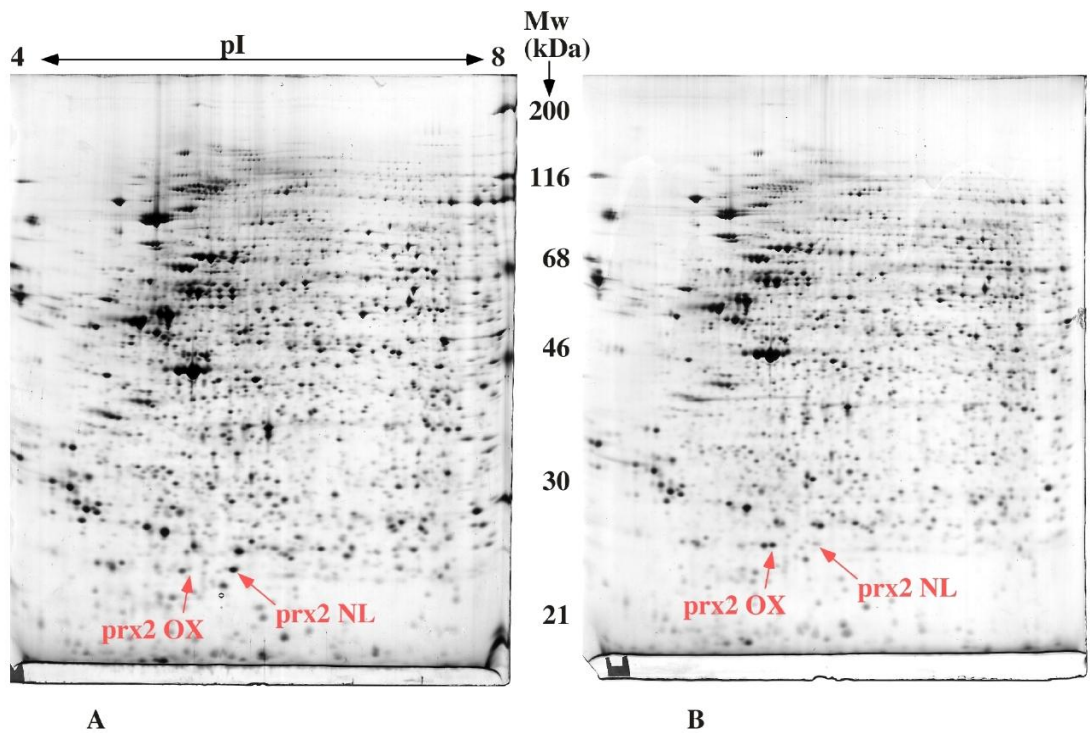

Figure 3: detection of peroxiredoxin modification
Jurkat cells were cultured either under control conditions (panel A) or stressed with 0.15mM t-Butylhydroperoxide for 2 hours before harvesting. Cells were lysed, and the extracts separated by two-dimensional electrophoresis: linear pH gradient 4-8, 10% acrylamide for the SDS gel, silver staining, 120 micrograms loaded on the first dimension.
The arrows indicate the position of the normal form of peroxiredoxin 2 (prx2NL) and of the oxidized form (prxOX). The change in pI (0.25 pH units) is due to the sole oxidation of the -SH group of the active site (cys51) into a sulfinic acid ($-SO_2H$), easily oxidized in the second dimension gel to a sulfonic acid ($-SO_3H$).